\documentstyle[aps,floats,12pt,graphicx]{revtex}

\begin{document}
\title{{\bf History and Physics of The Klein Paradox}}
\author{{\bf A\ Calogeracos}}
\author{NCA\ Research Consultants, PO Box 61147, Maroussi 151 22, Greece}
\author{(acal@hol.gr)}
\author{{\bf N Dombey}}
\author{Centre for Theoretical Physics, University of Sussex, Brighton BN1 9QJ,UK}
\author{(normand@sussex.ac.uk)}
\date{SUSX-TH-99-032}
\maketitle

\begin{abstract}
\newline
\noindent The early papers by Klein, Sauter and Hund which investigate
scattering off a high step potential in the context of the Dirac equation
are discussed to derive the 'paradox' first obtained by Klein. The
explanation of this effect in terms of electron-positron production is
reassessed. It is shown that a potential well or barrier in the Dirac
equation can become supercritical and emit  positrons or electrons 
spontaneously if the potential is strong enough. If the well or barrier 
is wide enough, a seemingly constant current is emitted. This 
phenomenon is transient whereas the tunnelling first calculated 
by Klein is time-independent. It is shown that tunnelling without 
exponential suppression occurs when an electron is incident on a 
high barrier, even when the barrier is not high enough to radiate. 
Klein tunnelling is therefore a property of relativistic wave 
equations and is not necessarily connected to particle emission. The 
Coulomb potential is investigated and it is shown that a heavy nucleus of
sufficiently large $Z$ will bind positrons. Correspondingly, 
as $Z$ increases the Coulomb barrier should become increasingly
transparent to positrons. This is an example of Klein tunnelling. Phenomena 
akin to supercritical positron emission may be studied experimentally 
in superfluid $^3$He
\end{abstract}

\section{Some History}

\subsection{Introduction to the Klein Paradox(es)}

\noindent Seventy years ago Klein \cite{klein} published a paper where he
calculated the reflection and transmission coefficients for electrons of
energy $E,$ mass $m$ and momentum $k$ incident on the potential step (Fig. 1)

\begin{equation}  \label{step}
V(x)=V,\;x>0;\,V(x)=0,\;x<0
\end{equation}

\noindent within the context of the new relativistic equation which had just
been published by Dirac\cite{dirac}. He found (see Section 2 below) that the
reflection and transmission coefficients $R_S,T_S$ if $V$ was large were
given by 
\begin{equation}
R_S=%
{\displaystyle {1-\kappa  \overwithdelims() 1+\kappa }}
^2\qquad T_S=\frac{4\kappa }{(1+\kappa )^2}  \label{Rs/Ts}
\end{equation}
\noindent where $\kappa $ is the kinematic factor 
\begin{equation}
\kappa =\frac pk\frac{E+m}{E+m-V}  \label{kappa}
\end{equation}
\noindent and $p$ is the momentum of the transmitted particle for $x>0.$ It
is easily seen from Eq. (\ref{kappa}) that when $E<V-m$, $\kappa $ seems to
be negative with the paradoxical result that the reflection coefficient $%
R_S>1$ while $T_S<0$. So more particles are reflected by the step than are
incident on it. This is what many articles and books call the Klein Paradox.
It is not, however, what Klein wrote down.

\smallskip\ 

\noindent Klein noted that Pauli had pointed out to him that for $x>0$, the
particle momentum is given by $p^2=(V-E)^2-m^2$ while the group velocity $%
v_g $ was given by

\begin{equation}  \label{group}
v_g=dE/dp=p/(E-V)
\end{equation}

\noindent So if the transmitted particle moved from left to right, $v_g$ was
positive implying that $p$ had to be assigned its negative value

\begin{equation}  \label{neg}
p=-\sqrt{(V-E)^2-m^2}
\end{equation}

\noindent With this choice of $p$

\begin{equation}
\kappa =\sqrt{\frac{(V-E+m)(E+m)}{(V-E-m)(E-m)}}  \label{kappa2}
\end{equation}

\noindent and $\kappa \geq 1$ ensuring that both $R_S$ and $T_S$ are
positive or zero and satisfy $R_S+T_S=1$ for $m$ $\leq E\leq V-m.$ Is there
still a paradox? The general consensus both now and for the authors who
followed Klein and did the calculation correctly is that there is. Let the
potential step $V\rightarrow \infty $ for fixed $E$ then from Eq. (\ref
{kappa2})$\,$ $\kappa $ tends to a finite limit and hence $T_S$ tends to a
non-zero limit. The physical essence of this paradox thus lies in the
prediction that according to the Dirac equation, fermions can pass through
strong repulsive potentials without the exponential damping expected in
quantum tunnelling processes. We have called this process Klein tunnelling 
\cite{kleinf}.

\smallskip\ 

\noindent We begin with a summary of the Dirac equation in one dimension in
the presence of a potential $V(x)$ and show how Klein's original result for $%
R_S$ and $T_S$ is obtained. We go on to the papers of Sauter in 1931, who
replaced Klein's potential step with a barrier with a finite slope, and then
to Hund in 1940 who realised that the Klein potential step gives rise to the
production of pairs of charged particles when the potential strength is
sufficiently strong. This result although not well known is a precursor of
the famous results of modern quantum field theory of Schwinger \cite{schwing}
and Hawking \cite{hawk} which show that particles are spontaneously produced
in the presence of strong electric and gravitational fields. In Part II we
turn to the underlying physics of the Klein paradox and show that particle
production and Klein tunnelling arise naturally in the Dirac equation: when
a potential well is deep enough it becomes supercritical (defined as the
potential strength for which the bound state energy $E=-m)$ and positrons
will be spontaneously produced. Supercriticality is well-understood \cite
{zeld}, \cite{grein2} and can occur in the Coulomb potential with finite
nuclear size when the nuclear charge $Z$ $>137$. Positron production via
this mechanism has been the subject of experimental investigations in heavy
ion collisions for many years. We then show that if a potential well is wide
enough, a steady but transient current will flow when the potential becomes
supercritical. In order to analyse these processes it is necessary to
introduce the concept of vacuum charge. We consider the implications of
these concepts for the Coulomb potential and for other physical phenomena
and we end by pointing out that Klein was unfortunate in that the example he
chose to calculate was pathological.

\subsection{The Dirac Equation in One Dimension}

\noindent In one-dimension it is unnecessary to use four-component Dirac
spinors. It is much easier to use two-component Pauli spinors instead \cite
{bruce}.We adopt the convention $\gamma _0=\sigma _z$, $\gamma _1=i\sigma _x$%
. The above choice agrees with $\gamma _i\gamma _j+\gamma _j\gamma
_i=2g_{ij}.$ The free Dirac Hamiltonian in one dimension is 
\[
H_0=-\sigma _yp+\sigma _zm 
\]
and so the Dirac equation takes the form 
\begin{equation}
(\sigma _x\frac \partial {\partial x}-E\sigma _z+m)\psi =0  \label{frd}
\end{equation}
In what follows ${\bf k}$ stands for the wavevector, $k$ for its magnitude
and $\varepsilon =\left| E\right| =+\sqrt{k^2+m^2}$. We try a plane wave of
the form 
\begin{equation}
\left( 
\begin{array}{c}
A \\ 
B
\end{array}
\right) e^{ikx-iEt}  \label{frs}
\end{equation}
and substitute in (\ref{frd}). The equation is satisfied by $A=ik,B=E-m$
where $E=\pm \varepsilon $. The positive energy (or particle) solutions have
the form 
\begin{equation}
N_{+}(\varepsilon )\left( 
\begin{array}{c}
ik \\ 
\varepsilon -m
\end{array}
\right) e^{ikx-i\varepsilon t}  \label{frp}
\end{equation}
and the negative energy (or hole) solutions are 
\begin{equation}
N_{-}(\varepsilon )\left( 
\begin{array}{c}
ik \\ 
-\varepsilon -m
\end{array}
\right) e^{ikx+i\varepsilon t}  \label{frn}
\end{equation}
where $N_{\pm }(\varepsilon )$ are appropriate normalization factors. If we
take the particle to be in a box of length 2$L$ with periodic boundary
conditions at $x=-L$ and $x=L$ we obtain

\begin{equation}
N_{+}(\varepsilon )=\frac 1{\sqrt{2L}\sqrt{2\varepsilon (\varepsilon -m)}}%
,N_{-}(\varepsilon )=\frac 1{\sqrt{2L}\sqrt{2\varepsilon (\varepsilon +m)}}
\label{norm}
\end{equation}

\noindent Alternatively we can use continuum states and energy
normalisation; then

\begin{equation}  \label{en}
N_{+}(\varepsilon )=\frac 1{\sqrt{2\pi }\sqrt{2\varepsilon (\varepsilon -m)}}%
,N_{-}(\varepsilon )=\frac 1{\sqrt{2\pi }\sqrt{2\varepsilon (\varepsilon +m)}%
}
\end{equation}

\subsection{The Klein Result}

\noindent In the presence of the Klein step, the Hamiltonian is 
\[
H_0=-\sigma _yp+V(x)+\sigma _zm
\]
where $V(x)$ is now given by Eq.(\ref{step}). The Dirac equation reads 
\begin{equation}
(\sigma _x\frac \partial {\partial x}-(E-V(x))\sigma _z+m)\psi =0
\label{eqn}
\end{equation}

\noindent Consider an electron incident from the left. The corresponding
wavefunction is 
\begin{equation}
\left( 
\begin{array}{c}
ik \\ 
E-m
\end{array}
\right) e^{ikx}+B\left( 
\begin{array}{c}
-ik \\ 
E-m
\end{array}
\right) e^{-ikx}  \label{sa}
\end{equation}
for $x<0,$ and 
\begin{equation}
F\left( 
\begin{array}{c}
-ip \\ 
V-E-m
\end{array}
\right) e^{-ipx}  \label{sc}
\end{equation}

\noindent for $x>0$ since that state is a hole state (see Fig. 2). It is
easy to see from Eqs. (\ref{sa}, \ref{sc}) that for continuity at $x=0$ we
require

\begin{eqnarray}
ik(1-B) &=&-ipF \\
(E-m)(1+B) &=&(V-E-m)F  \nonumber
\end{eqnarray}

\noindent giving

\[
\frac{1-B}{1+B}=\frac{-p}k\frac{E-m}{V-E-m}=\frac 1\kappa 
\]

\noindent in terms of the quantity $\kappa $ defined by Eq. (\ref{kappa}).
This gives the expression for $R_S=\left| B\right| ^2$ of Eq. (\ref{Rs/Ts})
above while that for $T_S$ follows from $R_S+T_S=1$.

\subsection{Sauter's Contribution}

\noindent Klein's surprising result was widely discussed by theoretical
physicists at the time. Bohr thought that the large transmission coefficient
that Klein found was because the Klein step was so abrupt. He discussed this
with Heisenberg and Sommerfeld and as a result Sommerfeld's assistant Sauter%
\cite{saut} in Munich calculated the transmission coefficient for a
potential of the form

\begin{equation}
V(x)=vx\,\qquad 0<x<L  \label{saut}
\end{equation}

\noindent with $V(x)=0$ for $x<0$ and $V(x)=vL$ for $x>L$ (Fig. 3). In order
to obtain negative energy states (holes) to propagate through the barrier as
in the Klein problem, we require $vL>2m$. Sauter's potential thus should
reduce to the Klein step if $v$ were very large. Sauter's potential is of
course more physical than Klein's: it simply represents a constant electric
field $E=-v$ in a finite region of space. Klein tunnelling in this case
would imply that low energy electrons could pass through a repulsive
constant electric field without exponential damping. Bohr conjectured that
the Klein result would only be reproduced if the Sauter field were so strong
that the potential difference $\Delta V>2m$ would be attained at distances
of the order of the Compton wavelength of the electron; that is to say that
the electric field strength $\left| E\right| =\left| v\right| >2m^2$.

\smallskip\ 

\noindent After a lengthy calculation involving the appropriate
hypergeometric functions, Sauter obtained the result he was seeking: he
obtained an expression for the reflection and transmission coefficients $R$
and $T$ which reduced to the Klein values $R_S$ and $T_S$ for $\left|
v\right| \sim m^2;$ nevertheless but for weaker fields he obtained

\begin{equation}
R\simeq 1\qquad T=e^{-\pi m^2/v}=e^{-(\pi m^2L/\Delta V)}  \label{sautres}
\end{equation}

\noindent a non-paradoxical result since it shows the
exponentially-suppressed tunnelling typical of quantum phenomena. What no
one realised at the time is that Sauter had anticipated Schwinger's\cite
{schwing} result of quantum electrodynamics by twenty years (see next
Section). Note also that Eq. (\ref{sautres}) shows explicitly that Bohr's
conjecture is correct: in order to violate the rule that tunnelling in
quantum mechanics is exponentially suppressed we require electric fields of
field strength $\left| E\right| =\left| v\right| \sim \pi m^2$.

\subsection{Hund's Contribution}

\noindent The next major contribution to the subject came ten years later.
Hund \cite{hund} looked again at the Klein step potential but from the
viewpoint of quantum field theory, not just the one particle Dirac equation.
He concentrated on charged scalar fields rather than spinor fields. He
considered both the Klein step potential and a sequence of step potentials.
His result was as surprising as Klein's original result. Hund found that
provided $\Delta V>2m$ where $\Delta V=V(\infty )-V(-\infty )$, then a
non-zero constant electric current $j$ had to be present where the current
was given by an integral over the transmission coefficient $T(E)$ with
respect to energy $E$. The current had to be interpreted as spontaneous
production out of the vacuum of a pair of oppositely charged particles. Hund
attempted to derive the same result for a spinor field but was unsuccessful:
it was left to Hansen and Ravndal \cite{hans} forty years later to
generalise this result to spinors (for a good discussion of the difference
between scalar and spinor fields incident on a Klein step see Manogue\cite
{mano}). We show in the Appendix for a Klein step or more general step
potential such as those considered by Hund and Sauter in the Dirac equation
that there is indeed a spontaneous current of electron-positron pairs
produced given by

\begin{equation}
\left\langle 0\right| j\left| 0\right\rangle =-\frac 1{2\pi }\int dET(E)
\label{pprod}
\end{equation}

\noindent in agreement with Hund's result for scalars. Eq. (\ref{pprod}) is
very powerful: it is a sort of optical theorem. If spontaneous pair
production occurs at a constant rate, then the time-independent reflection
and transmission coefficients must incorporate this process. If Sauter had
known of Eq. (\ref{pprod}), he would have been able to predict Schwinger's 
\cite{schwing} result on spontaneous pair production by a constant electric
field simply by using the value of the transmission coefficient he had
calculated in Eq. (\ref{sautres}).

\section{The Underlying Physics}

\subsection{Scattering by a Square Barrier}

\noindent We now investigate the underlying physics behind these phenomena.
Why is it that electrons can tunnel so easily through a high potential
barrier? Why are particles produced in strong potentials? Are these two
questions the same question; that is to say is the result that particles are
produced by a Klein step or other strong field the reason for Klein
tunnelling. To answer these questions we turn our attention to a potential
barrier which is not the Klein step but is similar and has better-defined
properties. This is the square barrier (Fig. 4)

\begin{equation}
V(x)=V,|\,x\,|<a;V(x)=0,|\,x\,|>a.  \label{barreq}
\end{equation}

\noindent Electrons incident from the left would not be expected to be able
to distinguish between a wide barrier (i.e. $ma>>1)$ and a Klein step. The
results are in fact not identical but they do display the same
characteristics.

\smallskip\ 

\noindent It is easy to show that the reflection and transmission
coefficients are given for a square barrier by \cite{jens} 
\begin{eqnarray}
R &=&\frac{(1-\kappa ^2)^2\sin ^2(2pa)}{4\kappa ^2+(1-\kappa ^2)^2\sin
^2(2pa)}\qquad  \label{barr} \\
T &=&\frac{4\kappa ^2}{4\kappa ^2+(1-\kappa ^2)^2\sin ^2(2pa)}
\end{eqnarray}

\noindent Note that tunnelling is easier for a barrier than a step: if

\begin{equation}  \label{quant}
2pa=N\pi
\end{equation}

\noindent corresponding to $E_N=V-\sqrt{m^2+N^2\pi ^2/4a^2}$ then the
electron passes right through the barrier with no reflection: this is called
a transmission resonance \cite{cdi}.

\smallskip\ 

\noindent As $a$ becomes very large for fixed $m,E$ and $V$, $pa$ becomes
very large and $\sin (pa)$ oscillates very rapidly. In those circumstances
we can average over the phase angle $pa$ using $\sin ^2(pa)=\cos ^2(pa)=%
\frac 12$ to find the limit

\begin{equation}  \label{inf}
R_\infty =\frac{(1-\kappa ^2)^2}{8\kappa ^2+(1-\kappa ^2)^2}\qquad T_\infty =%
\frac{8\kappa ^2}{8\kappa ^2+(1-\kappa ^2)^2}
\end{equation}

\noindent It may seem unphysical that $R_\infty $ and $T_\infty $ are not
the same as $R_S$ and $T_S$ but it is not: it is well known in
electromagnetic wave theory \cite{stratton} that reflection off a
transparent barrier of large but finite width (with 2 sides) is different
from reflection off a transparent step (with 1 side). The square barrier
thus demonstrates Klein tunnelling but it now arises in a more physical
problem than the Klein step. The zero of potential is properly defined for a
barrier whereas it is arbitrary for a step and the energy spectrum of a
barrier (which attracts positrons) or well (which attracts electrons) is
easily calculable. Particle emission from a barrier or well is described by
supercriticality: the condition when the ground state energy of the system
overlaps with the continuum ($E=m$ for a barrier; $E=-m$ for a well) and so
any connection between particle emission and the time-independent scattering
coefficients $R$ and $T$ can be investigated.

\subsection{Fermionic Emission from a Narrow Well}

\noindent We discussed the field theoretic treatment of this topic in a
previous paper \cite{cdi} which we refer to as CDI. We quickly review the
argument of that paper. Spontaneous fermionic emission is a non-static
process and in the case of a seemingly static potential, it is necessary to
ask how the potential was switched on from zero. We follow CDI in turning on
the potential adiabatically. We will consider the square well

\begin{equation}
V(x)=-V,|\,x\,|<a;V(x)=0,|\,x\,|>a  \label{well}
\end{equation}

\noindent but it is easiest to begin with the very narrow potential $%
V(x)=-\lambda \delta (x)$ which is the limit of a square well with $\lambda
=2Va$. The bound states are then very simple: for a given value of $\lambda $
there is just one bound state corresponding to either the even $(e)$ or odd (%
$o)$ wave functions \cite{cdi} with energy given by

\begin{equation}  \label{delta}
E=m\cos \lambda \quad (e)\qquad E=-m\cos \lambda \quad (o)
\end{equation}

\noindent When the potential is initially turned on and $\lambda $ is small
the bound state is even and its energy $E$ is just below $E=m$. As $\lambda $
increases, $E$ decreases and at $\lambda =\pi /2$ $,$ $E$ reaches zero. For $%
\lambda >\pi /2$, $E$ becomes negative. Assuming that we started in the
vacuum state and therefore that the well was originally vacant, we now have
for $\lambda >\pi /2$ the absence of a negative energy state which must be
interpreted as the presence of a (bound) positron according to Dirac's hole
theory. Let $\lambda $ increase further and $E$ decreases further until at $%
\lambda =\pi ,$ $E=-m$ which is the supercriticality condition. So for $%
\lambda >\pi ,$ the bound positron acquires sufficient energy to escape from
the well. This is the phenomenon of spontaneous positron production as
described originally by Gershtein and Zeldovich \cite{zeld} and Pieper and
Greiner \cite{grein2}. Note that this picture requires that positrons (as
well as electrons) are bound by potential wells when the potential strength
is large enough: we return to this point later when we discuss the Coulomb
potential.

\subsection{Digression on Vacuum Charge}

\noindent How is it possible to conserve charge and produce positrons out of
the vacuum? This question has been a fruitful ground for theorists in recent
years. The key point is that the definition of the vacuum state of the
system (and of the other states) depends on the background potential: this
leads to the concept of vacuum charge \cite{stone}, \cite{blank}. At this
point a single particle interpretation of a potential in the Dirac equation
is insufficient and field theory becomes necessary (as is also seen in the
discussion of radiation from the Klein step in the Appendix). But
nevertheless it turns out that once the concept of vacuum charge is
introduced, first quantisation is all that is necessary to determine its
value. We shall refer the reader to CDI for a proper treatment of vacuum
charge; we just write down the essential equations here.

\smallskip\ 

\noindent The total charge is defined by (according to our conventions the
electron charge is $-$1)\ 
\begin{equation}
Q(t)=\int dx\rho (x,t)=-\frac 12\int dx\left[ \psi ^{\dagger }(x,t),\psi
(x,t)\right]  \label{ch1}
\end{equation}
Writing the wave function $\psi (x,t)$ in terms of creation and annihilation
operators we eventually find that 
\begin{equation}
Q=Q_p+Q_0  \label{ch}
\end{equation}
where the particle charge $Q_p$ is an operator which counts the number of
electrons in a state minus the number of positrons while the vacuum charge $%
Q_0$ is just a number which is defined by the difference in the number of
positive energy and negative energy states of the system: 
\begin{equation}
Q_0=\frac 12\left\{ \sum_k(\text{states with }E>0)-\sum_k(\text{states with }%
E<0)\right\}  \label{qo}
\end{equation}
\noindent Given the definition of the vacuum we immediately get 
\begin{equation}
\left\langle 0\right| Q\left| 0\right\rangle =Q_0  \label{vgch}
\end{equation}

\noindent We illustrate the use of the vacuum charge by returning to the
delta function potential $V(x)=-\lambda \delta (x)$. For $\lambda $ just
larger than $\pi /2$, $Q_p=+1$ because a positron has been created, but now
the vacuum charge $Q_0=-1$ because the number of positive energy states has
decreased by one while the number of negative energy states has increased by
one. So the total charge $Q$ is in fact conserved. As the potential is
increased further, $\lambda $ will reach $\pi .$ where $E=-m$ and the bound
positron reaches the continuum and becomes free. Note that at
supercriticality, there is no change in vacuum charge; the change occurs
when $E$ crosses the zero of energy. Note also that at supercriticality the
even bound state disappears and the first odd state appears.

\smallskip\ 

\noindent We can continue to increase $\lambda $ and count positrons: the
total number of positrons produced for a given $\lambda $ is the number of
times $E$ has crossed $E=0;$ that is 
\begin{equation}
Q_p=Int\,[\frac \lambda \pi +\frac 12]  \label{charge0}
\end{equation}

\noindent and $Q_0=$ -$Q_p$ where $Int[x]$ denotes the integer part of $x.$
For positron emission the more interesting quantity is the number of
supercritical positrons $Q_S$, that is the number of states which have
crossed $E=-m$. This is given by

\begin{equation}  \label{super}
Q_S=Int\,[\frac \lambda \pi ]
\end{equation}

\subsection{Wide Well}

\noindent We can now return to the case that we are interested in which is
that of a wide well or barrier. So let us consider the general case of a
square well potential of strength $V>2m$ and then look at a wide well for
which $ma>>1$ most closely corresponding to the Klein step. We follow the
discussion given in our papers CDI and CD \cite{kleinf}. We must find first
the condition for supercriticality and then the number of bound and
supercritical positrons produced for a given $V.$

\smallskip\ 

\noindent The bound state spectrum for the well $V(x)=-V,|\,x\,|<a;V(x)=0,|%
\,x\,|>a$ is easily obtained: there are even and odd solutions given by the
equations

\begin{equation}
\tan pa=\sqrt{\frac{(m-E)(E+V+m)}{(m+E)(E+V-m)}}  \label{even}
\end{equation}

\begin{equation}
\tan pa=-\sqrt{\frac{(m+E)(E+V+m)}{(m-E)(E+V-m)}}  \label{odd}
\end{equation}

\noindent where now the well momentum is given by $p^2=(E+V)^2-m^2$. We have
changed the sign of $V$ so that it is now attractive to electrons rather
than positrons in order to conform with other authors who have studied
supercritical positron emission rather than electron emission .

\smallskip\ 

\noindent 

From Eq (\ref{even}) we see that the ground state becomes supercritical when 
$pa=\pi /2$ and therefore $V_1^c=m+\sqrt{m^2+\pi ^2/4a^2}.$ From Eq (\ref
{odd}) the first odd state becomes supercritical when $pa=\pi $ and $V_2^c=m+%
\sqrt{m^2+\pi ^2/a^2}.$ Clearly the supercritical potential corresponding to
the Nth positron is

\begin{equation}  \label{Nth}
V_N^c=m+\sqrt{m^2+N^2\pi ^2/4a^2}
\end{equation}

\noindent It follows from Eq (\ref{Nth}) that $V=2m$ is an accumulation
point of supercritical states as $ma\rightarrow \infty $. Furthermore it is
a threshold: a potential $V$ is subcritical if $V<2m$. It is not difficult
to show for a given $V>2m$ that the number of supercritical positrons is
given by

\begin{equation}  \label{super2}
Q_S=Int[(2a/\pi )\sqrt{V^2-2mV}]
\end{equation}

\noindent The corresponding value of the total positron charge $Q_p$ can be
shown using Eqs (\ref{even},\ref{odd}) to satisfy

\begin{equation}  \label{charge}
Q_p-1\leq Int[(2a/\pi )\sqrt{V^2-m^2}]\leq Q_p
\end{equation}

\noindent so for large $a$ we have the estimates

\begin{equation}  \label{est}
Q_p\sim (2a/\pi )\sqrt{V^2-m^2};\quad Q_S\sim (2a/\pi )\sqrt{V^2-2mV}
\end{equation}
$\qquad $

\noindent Now we can build up an overall picture of the wide square well $%
ma>>1$. When $V$ is turned on from zero in the vacuum state an enormous
number of bound states is produced. As $V$ crosses $m$ a very large number $%
Q_p$ of these states cross $E=0$ and become bound positrons. As $V$ crosses $%
2m$ a large number $Q_S$ of bound states become supercritical together. This
therefore gives rise to a positively charged current flowing from the well.
But in this case, unlike that of the Klein step, the charge in the well is
finite and therefore the particle emission process has a finite lifetime.
Nevertheless, for $ma$ large enough the transient positron current for a
wide barrier is approximately constant in time for a considerable time as we
shall see in the next section.\ 

\subsection{Emission Dynamics}

\noindent We now restrict ourselves to the case $V=2m+\Delta $ with $\Delta
<<m$. This is not necessary but it avoids having to calculate the dynamics
of positron emission while the potential is still increasing beyond the
critical value. We can assume all the positrons are produced almost
instantaneously as the potential passes through $V=2m.$ It also means that
the kinematics are non-relativistic. Hence for a sufficiently wide well so
that $\Delta a$ is large, $Q_S\sim (2a/\pi )\sqrt{2m\Delta }$. The well
momentum of the Nth supercritical positron is still given by Eq (\ref{quant}%
) $p_Na=N\pi /2$ which corresponds to an emitted positron energy $%
|\,E_N\,|=2m+\Delta -\sqrt{p_N^2+m^2}>m$. Note that the emitted energies
have discrete values although for $a$ large, they are closely spaced.

\smallskip\ 

\noindent The lifetime $\tau $ of the supercritical well is given by the
time for the slowest positron to get out of the well. The slowest positron
is the deepest lying state with $N=1$ and momentum $p_1=\pi /2a$. Hence $%
\tau \approx ma/p_1=2ma^2/\pi .$ So the lifetime is finite but scales as $%
a^2 $. But a large number of positrons will have escaped well before $\tau $%
. There are $Q_S$ supercritical positrons initially and their average
momentum $\overline{p}$ corresponds to $N=Q_S/2$; hence $\overline{p}=\sqrt{%
m\Delta /2}$ which is independent of $a$. Thus a transient current of
positrons is produced which is effectively constant in time for a long time
of order $\overline{\tau }$ $=a\sqrt{2m/\Delta }$. We thus see that the
square well (or barrier) for $a$ sufficiently large behaves just like the
Klein step: it emits a seemingly constant current with a seemingly
continuous energy spectrum. But initially the current must build up from
zero and eventually must return to zero. So the well/barrier is a
time-dependent physical entity with a finite but long lifetime for emission
of supercritical positrons or electrons.

\smallskip\ 

\noindent Note again that the transmission resonances of the
time-independent scattering problem coincide with the energies of particles
emitted by the well or barrier. It is therefore tempting to use the Pauli
principle to explain the connection. Following Hansen and Ravndal \cite{hans}%
, we could say that $R$ must be zero at the resonance energy because the
electron state is already filled by the emitted electron with that energy.
But it is easy to show that the reflection coefficient is zero for bosons as
well as fermions of that energy, and no Pauli principle can work in that
case. Furthermore emission ceases after time $\tau $ whereas $R=0$ for times 
$t>\tau $ . It follows that we must conclude that Klein tunnelling is a
physical phenomenon in its own right, independent of any emission process.
It seems that Klein tunnelling is indeed distinct from the particle emission
process: to show this is so we return to the square barrier to show that
Klein tunnelling occurs even when the barrier is subcritical.

\subsection{Klein Tunnelling and the Coulomb Barrier}

\noindent It is clear from Eq (\ref{barr}) that while the reflection
coefficient $R$ for a square barrier cannot be $0$, neither is the
transmission coefficient $T$ exponentially small for energies $E<V$ when $%
V>2m$ even though the scattering is classically forbidden. The simplest way
to understand this is to consider the negative energy states under the
potential barrier as corresponding to physical particles which can carry
energy in exactly the same way that positrons are described by negative
energy states which can carry energy. It follows from Eq (\ref{Rs/Ts}) that $%
R_S$ and $T_S$ correspond to reflection and transmission coefficients in
transparent media with differing refractive indices: thus $\kappa $ is
nothing more than an effective fermionic refractive index corresponding to
the differing velocities of propagation by particles in the presence and
absence of the potential. On this basis, tuning the momentum $p$ to obtain a
transmission resonance for scattering off a square barrier is nothing more
than finding the frequency for which a given slab of refractive material is
tranparent. This is not a new idea. In Jensen's words ''A potential hill of
sufficient height acts as a Fabry-Perot etalon for electrons, being
completely transparent for some wavelengths, partly or completely reflecting
for others'' \cite{bak}.

\smallskip\ 

\noindent We can now look in more detail at Klein tunnelling: both in terms
of our model square well/barrier problem and at the analogous Coulomb
problem. The interesting region is where the potential is strong but
subcritical so that emission dynamics play no role and sensible time
independent scattering parameters can be defined. For electron scattering
off the square barrier $V(x)=V$ we would thus require $V<$ $V_1^c=m+\sqrt{%
m^2+\pi ^2/4a^2}$ together with $V>2m$ so that positrons can propagate under
the barrier. For the corresponding square well $V(x)=-V$ there are negative
energy bound states $0>E>-m$ provided that $V>$ $\sqrt{m^2+\pi ^2/4a^2}$
[cf. Eq.(\ref{charge})]. So when the potential well is deep enough, it will
in fact bind positrons. Correspondingly, a high barrier will bind electrons.
It is thus not surprising that electrons can tunnel through the barrier for
strong subcritical potentials since they are attracted by those potential
barriers. Another way of seeing this phenomenon is by using the concept of
effective potential $V_{eff}(x)$ which is the potential which can be used in
a Schrodinger equation to simulate the properties of a relativistic wave
equation. For a potential $V(x)$ introduced as the time-component of a
four-vector into a relativistic wave equation (Klein-Gordon or Dirac), it is
easy to see that $2mV_{eff}(x)=2EV(x)-V^2(x).$ Hence as the energy $E$
changes sign, the effective potential can change from repulsive to
attractive.

\smallskip\ 

\noindent For the pure Coulomb potential, it is well known that there is
exponential suppression of the wave functions for a repulsive potential
compared with an attractive potential. For example, if $\rho =\left| \psi
(0)\right| _{pos}^2/\left| \psi (0)\right| _{el}^2$ is the ratio of the
probability of a positron penetrating a Coulomb barrier to reach the origin
compared with the probability of an electron of the same energy, then if the
particles are non-relativistic 
\begin{equation}
\rho =e^{-2\pi Z\alpha E/p}  \label{nonrel}
\end{equation}
\noindent where p and $E$ are the particle momenta and energies and this is
exponentially small as $p\rightarrow 0$ \cite{LL}. But if the particles are
relativistic \cite{rose} 
\begin{equation}
\rho =fe^{-2\pi Z\alpha }  \label{rel}
\end{equation}
\noindent where $f$ is a ratio of complex gamma-functions and is
approximately unity for large $Z$ . So $\rho \sim e^{-2\pi Z\alpha }\approx
10^{-3}$ for $Z\alpha \sim 1$ which is not specially small although it still
decreases exponentially with $Z..$

\smallskip\ 

\noindent In order to demonstrate Klein tunnelling for a Coulomb potential
we require first the inclusion of nuclear size effects so that the potential
is not singular at $r=0$ and second that $Z$ is large enough so that bound
positron states are present. This means that $Z$ must be below its
supercritical value $Z_c$ of around $170$ but large enough for the $1s$
state to have $E<0$. The calculations of references \cite{zeld} and \cite
{grein2} which depend on particular models of the nuclear charge
distribution give this region as $150<Z<Z_c$ which unfortunately will be
difficult to demonstrate experimentally. Nevertheless, the theory seems to
be clear: in this subcritical region positrons should no longer obey a
tunnelling relation which decreases exponentially with Z such as that of Eq.
(\ref{rel}). Instead the Coulomb barrier should become more transparent as $%
Z $ increases, at least for low energies. By analogy with the square barrier
we may expect that maximal transmission for positron scattering on a Coulomb
potential should occur around $Z=Z_c$ although the onset of supercriticality
implies that time independent scattering quantities may no longer be
well-defined. We are now carrying out further detailed calculations to
clarify the situation for positron scattering off nuclei with $Z$ near $Z_c$
to see if we can simulate Klein tunnelling.

\section{Conclusions}

\noindent It seems that Klein was very unfortunate in that the potential
step he considered is pathological and therefore a misleading guide to the
underlying physics. Klein's step represents a limit in which time-dependent
emission processes become time-independent and therefore a relationship
between the emitted current and the transmission coefficient exists, as we
show in the Appendix. In general no direct relationship would exist between
the transient current emitted and the time-independent transmission
coefficient. The physics of the Dirac equation which underlies Klein's
result is rich: it includes spontaneous fermionic production by strong
potentials and the separate phenomenon of Klein tunnelling by means of the
negative energy states characteristic of relativistic wave equations,
similar to interband tunnelling in semiconductors \cite{semi}. Spontaneous
positron production due to supercriticality has not yet been unambiguously
demonstrated experimentally in heavy ion collisions but experiments on
superfluid $^3$He-B\cite{lanc1}, \cite{lanc2} have displayed anomalous
effects when the velocity of a body moving in the fluid exceeds the critical
Landau velocity $v_L$. These experiments have now been interpreted in the
same way as supercritical positron production\cite{cv}. It may well be that
fermionic many-body systems can be used to demonstrate the fundamental
quantum processes which Klein unearthed seventy years ago

\smallskip\ 

\noindent We wo uld like to thank A. Anselm, G Barton, J D Bjorken, B.
Garraway, R Hall, L B Okun, R Laughlin, G E Volovik and D Waxman for advice
and help.

\section{Appendix: Pair Production by a Step Potential}

\noindent Consider the Klein step of Eq. (\ref{step}) for $V>2m$. We will
show that the expectation value of the current in the vacuum state in the
presence of the step is non-zero which means that the Klein step produces
electron-positron pairs out of the vacuum at a constant rate. The derivation
hinges on a careful definition of the vacuum state.We use the derivation of
CD2\cite{kleinf}.

\subsection{The normal modes in the presence of the Klein step.}

\noindent An energy-normalised positive energy or particle solution to the
Dirac equation can be written from eq. (\ref{en})

\begin{equation}  \label{pos}
\sqrt{%
{\displaystyle {\varepsilon +m \over 2k}}
}\left( 
\begin{array}{c}
i \\ 
{\displaystyle {k \over E+m}}
\end{array}
\right) e^{ikx}
\end{equation}

\noindent A negative energy or hole solution reads 
\[
\sqrt{%
{\displaystyle {\varepsilon -m \over 2k}}
}\left( 
\begin{array}{c}
i \\ 
{\displaystyle {k \over E+m}}
\end{array}
\right) e^{ikx} 
\]

\noindent Scattering is usually described by a solution describing a wave
incident (say from the left) plus a reflected wave (from the right) plus a
transmitted wave (to the right). It is convenient here to use waves of
different form either describing a wave (subscript $L)$ incident from the
left with no reflected wave or describing a wave (subscript $R$) incident
from the right with no reflected wave. Particle and hole wavefunctions will
be denoted by $u$ and $v$ respectively. It is clear that the nontrivial
result we are seeking arises from the overlap of the hole continuum $E<V-m$
on the right with the particle continuum $E>m$ on the left. We are thus
concerned with wavefunctions with energies in the range $m<E<V-m.$ The
expressions for $u_L,u_R$ in this energy range are given below.

\ 

\begin{equation}
\begin{array}{c}
\sqrt{2\pi }u_L(E,x)=%
{\displaystyle {\sqrt{2\kappa } \over \kappa +1}}
\sqrt{%
{\displaystyle {E+m \over k}}
}\left( 
\begin{array}{c}
i \\ 
{\displaystyle {k \over E+m}}
\end{array}
\right) e^{ikx}\theta (-x)+ \\ 
\\ 
\left\{ 
{\displaystyle {\kappa -1 \over \kappa +1}}
\sqrt{%
{\displaystyle {V-E-m \over 2\left| p\right| }}
}\left( 
\begin{array}{c}
i \\ 
{\displaystyle {\left| p\right|  \over E+m-V}}
\end{array}
\right) e^{i\left| p\right| x}+\sqrt{%
{\displaystyle {V-E-m \over 2\left| p\right| }}
}\left( 
\begin{array}{c}
i \\ 
{\displaystyle {-\left| p\right|  \over E+m-V}}
\end{array}
\right) e^{-i\left| p\right| x}\right\} \theta (x)
\end{array}
\label{w1}
\end{equation}

\begin{equation}
\begin{array}{c}
\sqrt{2\pi }u_R(E,x)=\left\{ 
{\displaystyle {1-\kappa  \over 1+\kappa }}
\sqrt{%
{\displaystyle {E+m \over 2k}}
}\left( 
\begin{array}{c}
i \\ 
{\displaystyle {k \over E+m}}
\end{array}
\right) e^{ikx}+\sqrt{%
{\displaystyle {E+m \over 2k}}
}\left( 
\begin{array}{c}
i \\ 
{\displaystyle {-k \over E+m}}
\end{array}
\right) e^{-ikx}\right\} \theta (-x)+ \\ 
\\ 
+%
{\displaystyle {\sqrt{2\kappa } \over \kappa +1}}
\sqrt{%
{\displaystyle {V-E-m \over \left| p\right| }}
}\left( 
\begin{array}{c}
i \\ 
{\displaystyle {\left| p\right|  \over E+m-V}}
\end{array}
\right) e^{i\left| p\right| x}\theta (x)
\end{array}
\label{w2}
\end{equation}

\noindent We write $\left| p\right| $ rather than $p$ in these equations
since the group velocity is negative for $x>0$ (cf. Eq. (\ref{sc}))

\smallskip\ 

\noindent We need to evaluate the currents corresponding to the solutions of
Eqs (\ref{w1},\ref{w2}). According to our conventions $\alpha _x\,=\gamma
_0\gamma _x=-\sigma _y$ so 
\begin{equation}
j_L\equiv -u_L^{\dagger }(E,x)\sigma _yu_L(E,x)=-\frac{2\kappa /\pi }{%
(\kappa +1)^2}  \label{c1}
\end{equation}
\begin{equation}
j_R\equiv -u_R^{\dagger }(E,x)\sigma _yu_R(E,x)=-\frac{2\kappa /\pi }{%
(\kappa +1)^2}  \label{c2}
\end{equation}

\subsection{The definition of the vacuum and the vacuum expectation value of
the current.}

Now expand the wave function $\psi $ in terms of creation and annihilation
operators which refer to our left- and right-travelling solutions: 
\begin{equation}
\begin{array}{c}
\psi (x,t)=\int dE\{a_L(E)u_L(E,x)e^{-iEt}+a_R(E)u_R(E,x)e^{-iEt}+ \\ 
+b_L^{\dagger }(E)v_L(E,x)e^{iEt}+b_R^{\dagger }(E)v_R(E,x)e^{iEt}\}
\end{array}
\label{exp}
\end{equation}

\noindent with $\psi ^{\dagger }$ given by the Hermitian conjugate
expansion.We must now determine the appropriate vacuum state in the presence
of the step. States described by wavefunctions $u_L(E,x)$ and $v_L(E,x)$
correspond to (positive energy) electrons and positrons respectively coming
from the left. Hence with respect to an observer to the left (of the step)
such states should be absent from the vacuum state, so

\begin{equation}  \label{a1}
a_L(E)\left| 0\right\rangle =0,\,b_L(E)\left| 0\right\rangle =0
\end{equation}

\noindent Wavefunctions $u_R(E,x)$ for $E>m+V$ describe for an observer to
the right, electrons incident from the right. These are not present in the
vacuum state hence 
\begin{equation}  \label{a2}
a_R(E)\left| 0\right\rangle =0\text{ for }E>m+V
\end{equation}

\noindent Wavefunctions $v_R(E,x)$ describe, again with respect to an
observer to the right, positrons incident from the right; again 
\begin{equation}  \label{b2}
b_R(E)\left| 0\right\rangle =0\text{ }
\end{equation}

\noindent The wavefunctions that play the crucial role in the Klein problem
belong to the set $u_R(E,x)$ for $m<E<V-m.$ For an observer to the right
these states are positive energy positrons and hence they should be filled
in the vacuum state, i.e.

\begin{equation}  \label{vac}
a_R^{\dagger }(E)a_R(E^{\prime })\left| 0\right\rangle =\delta (E-E^{\prime
})\left| 0\right\rangle \text{ , }m<E<V-m
\end{equation}

\noindent Having specified the vacuum the next and final step is the
calculation of the vacuum expectation value \ of the current: 
\begin{equation}  \label{vev}
\left\langle 0\right| j\left| 0\right\rangle =\frac 12\left( -\left\langle
0\right| \psi ^{\dagger }\sigma _y\psi \left| 0\right\rangle +\left\langle
0\right| \psi \sigma _y\psi ^{\dagger }\left| 0\right\rangle \right)
\end{equation}

\noindent Substituting (\ref{exp}) in (\ref{vev}) and noticing that all
terms involving $v_L$ and $v_R$ can be dropped since the corresponding
energies lie outside the interesting range $m<E<V-m$ we end up with 
\begin{equation}  \label{sum}
\begin{array}{c}
\left\langle 0\right| j\left| 0\right\rangle =-\frac 12\int dEdE^{\prime
}\{\left\langle 0\right| a_L^{\dagger }(E)a_L(E^{\prime })\left|
0\right\rangle u_L^{\dagger }(E,x)\sigma _yu_L(E^{\prime },x)+ \\ 
\\ 
+\left\langle 0\right| a_L(E)a_L^{\dagger }(E^{\prime })\left|
0\right\rangle u_L^{\dagger }(E^{\prime },x)\sigma _yu_L(E,x)-\left\langle
0\right| a_R^{\dagger }(E)a_R(E^{\prime })\left| 0\right\rangle u_R^{\dagger
}(E,x)\sigma _yu_R(E^{\prime },x)+ \\ 
\\ 
+\left\langle 0\right| a_R(E)a_R^{\dagger }(E^{\prime })\left|
0\right\rangle u_R^{\dagger }(E^{\prime },x)\sigma _yu_R(E,x)\}
\end{array}
\end{equation}

\smallskip\ 

\noindent The first term in (\ref{sum}) vanishes due to (\ref{a1}). The
second term becomes

$u_L^{\dagger }(E^{\prime },x)\sigma _yu_L(E,x)\delta (E-E^{\prime })$ if we
use the anticommutation relations and (\ref{a1}). The third term yields $%
-u_R^{\dagger }(E,x)\sigma _yu_R(E,x)\delta (E-E^{\prime })$ using (\ref{vac}%
) and the fourth term vanishes using the anticommutation relations (i.e. the
exclusion principle; the state $\left| 0\right\rangle $ already contains an
electron in the state $u_R$ hence we get zero when we operate on it with $%
a_R^{\dagger }$). One energy integration is performed immediately using the $%
\delta $ function. We obtain

\begin{equation}
\left\langle 0\right| j\left| 0\right\rangle =\frac 12\int dE(-j_L+j_R)=-%
\frac 1{2\pi }\int dE\frac{4\kappa (E)}{(\kappa (E)+1)^2}=-\frac 1{2\pi }%
\int dET_S(E)  \label{res}
\end{equation}

\noindent where the energy integration is over the Klein range $m<E<V-m$.

\smallskip\ 

\noindent It is now straightforward to generalise Eq (\ref{res}) to any step
potential for which $V(x<0)=V_1;\,V(x>L)=V_2$ and $V_2-V_1>2m$ such as those
considered by Sauter and Hund to obtain Eq (\ref{pprod} linking the pair
production current with the transmission coefficient.

\begin{figure}[bthp]
\begin{center}\leavevmode
\includegraphics[width=0.7\linewidth]{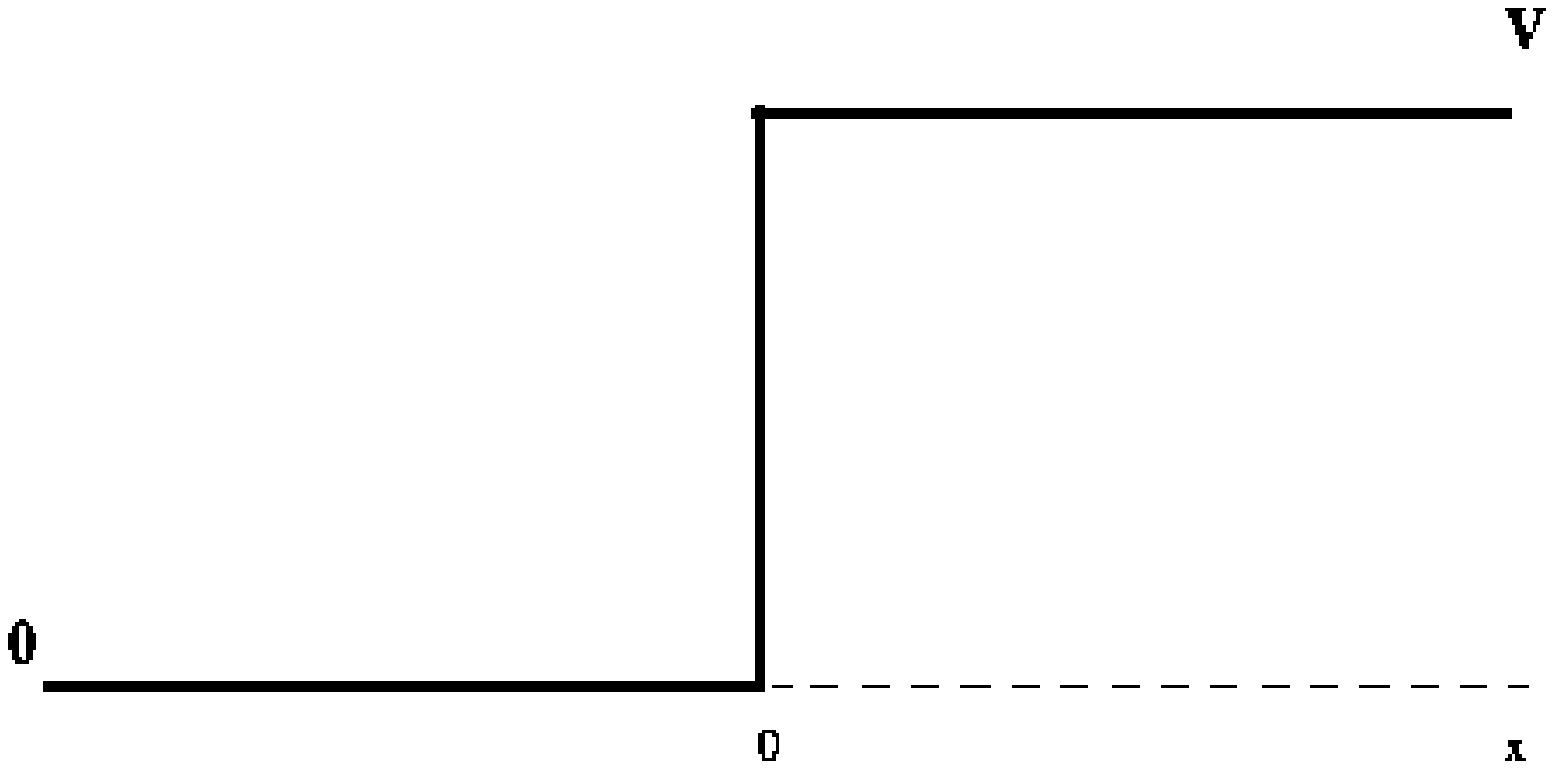}
\bigskip
\caption{The potential $V(x)$ of the Klein step}  
\end{center}
\end{figure}
\begin{figure}[bthp]
\begin{center}\leavevmode
\includegraphics[width=0.7\linewidth]{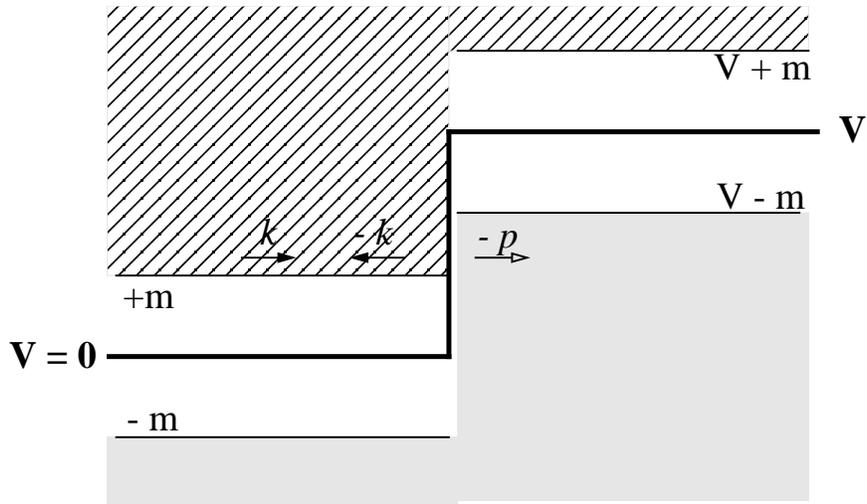}
\bigskip
\caption{An electron of energy $E$ scattering off a Klein step of
height $V>2m$. The electrons are shown with solid arrowheads; the hole state
has a hollow arrowhead. The particle continuum (slant bacground) and the
hole continuum (shaded background) overlap when $m<E<V-m.$}  
\end{center}
\end{figure}
\begin{figure}[bthp]
\begin{center}\leavevmode
\includegraphics[width=0.7\linewidth]{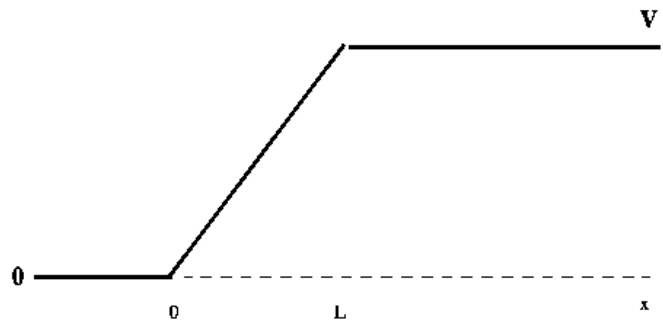}
\bigskip
\caption{A potential $V(x)$ of the Sauter form representing constant electric
field in the region $0<x<L$.}  
\end{center}
\end{figure}
\begin{figure}[bthp]
\begin{center}\leavevmode
\includegraphics[width=0.7\linewidth]{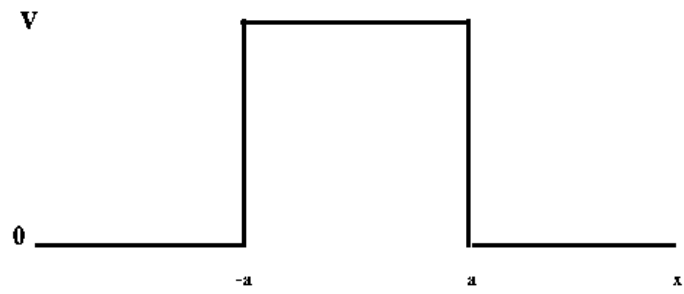}
\bigskip
\caption{A potential $V(x)$ representing a square barrier of height $V$
in the region $-a<x<a$.}  
\end{center}
\end{figure}
\end{document}